\documentclass[12pt]{iopart}

\usepackage{graphicx}
\usepackage{bm}

\begin{document}

\title{Dineutron clusters in $^7$He and $^8$He structure}

\author{E Garrido$^1$ and A S Jensen$^2$}

\address{$^1$Instituto de Estructura de la Materia, CSIC, Serrano 123, E-28006 Madrid, Spain}
\address{$^{2}$Department of Physics and Astronomy, Aarhus University, DK-8000 Aarhus C, Denmark}
\ead{e.garrido@csic.es}
\vspace{10pt}
 %\begin{indented}
%\item[]August 2017 (minor update March 2024)
%\end{indented}

\begin{abstract}
The hyper-radial barrier strongly hinders formation of more than three
clusters.  We investigate how well the dominating cluster components
in $^7$He and $^8$He, respectively can be described as
$\alpha$+$n$+$^2n$ and $\alpha$+$^2n$+$^2n$, where $^2n$ is the
dineutron.  Effective interactions compatible with $^5$He and $^6$He
are used.  We vary the lesser known $n$-$^2n$ and $^2n$-$^2n$
interactions, where very small strengths are required.  We provide
energies, radii, and partial wave decomposition of all computed,
predicted or measured, ground and resonance states.  We predict
substructures within each of the three-body quantum states.  We also
calculate the neutron-structure sensitive invariant mass spectrum of
the four-nucleon system, after fast removal of the $\alpha$-particle
from $^8$He.  We show that all available experimental information are
fairly well reproduced.  Very little room is left for variation of the
effective interaction parameters. Thus, the dominating features of the
subsequently derived reaction and structure properties are well
supported.
\end{abstract}

%\maketitle

\section{Introduction}

Cluster structures have been recognized in a number of nuclear systems
\cite{fre07}.  The description of such structures by means of ab-initio
calculations are exceedingly difficult, since the bare interactions and the 
intrinsic cluster structures would be involved, and the  simple cluster picture 
would be lost.  The necessity of few-body calculations with effective
interactions is then obvious.  

Concerning weakly bound many-neutron systems, which structures would characterize them?
A general observation formulated as a theorem is helpful: ``Two and
three clusters are strongly preferred'' \cite{jen04,rii13}.
Therefore, four (and more)-body halos do not exist. The proof of this
statement is that the many-body hyper-radial potential barrier
prevents more than three clusters.  The bulk of the wave function
would be either squeezed inside the barrier and overlap with the core
or be pushed outside the barrier without any chance to bind even in a
resonance state.  The intrinsic structures of the clusters do not have
to be point-like, although they are treated as one entity.  The
effective interactions between clusters should be adjusted to account
for this.  The accuracy is reduced, if more than one three-cluster
division is contributing.

\subsection{Neutrons around a core}

A tightly bound core surrounded by one or two weakly bound
neutrons are prototypes of two- and three-body clusters as
exemplified by, respectively, $^{11}$Be($^{10}$Be$+n)$ \cite{esb95}, and
$^6$He($\alpha+n+n$) or $^{11}$Li($^{10}$Li$+n+n$) \cite{zhu93,mor07}.
The neutron-neutron probability distribution for these systems has a
peak at short distance, which is characterized by a relative $s$-wave and
total spin equal to zero.  The core is spatially well separated.

Although a cluster as an object making part of a bigger structure is
typically understood as a bound constituent of the big system, unbound
entities as constituents in clusterized systems do actually become
possible when another body indirectly delivers a sufficient
attraction.  The unbound dineutron, $^2n$, can subsequently be
considered as a constituent particle, often described as a spatially
correlated structure of two neutrons. The structure is not unlike
models of two neutrons in a shell of given orbital and total angular
momentum.  In shell models, they have a strong tendency to form $0^+$
spin-correlated nucleon pairs arranged in shells outside a core.

Dineutron correlation has previously been proposed as structures of
neutron dripline nuclei, e.g. in the halo defining the description of
$^{11}$Li($^{9}$Li+$^2n$) \cite{han86}, or,  more recently, in three-body
calculations like $^8$Li($\alpha$+deuteron+$^2n$) \cite{gar25},
$^7$H(triton+$^2n$+$^2n$) \cite{aoy09},
$^8$He($\alpha$+$^2n$+$^2n$) \cite{kan07,kob13,yam23,laz23,nak25}, and
perhaps even $^{28}$O($^{24}$O+$^2n$+$^2n$) \cite{kon23}.  Experiments
on these systems are adding information \cite{yam23,nak25,due22,yan23}
to be explained by theory.  The loosely bound dineutrons are part of a
cluster structure, where core and valence constituents (neutrons and
dineutrons) essentially are completely decoupled.

From this, we deduce that the dominating structures of
$^8$He($\alpha$+$^2n$+$^2n$) and $^7$He($\alpha$+$n$+$^2n$), are the indicated
three-body clusters.  This can be viewed as a continuation of the
structure of the lighter Helium isotopes.  The $\alpha$-particle,
$^4$He, surrounded itself with an increasing number of neutrons.
First, the different partial-wave $\alpha$-$n$ interactions must match
the description of $^5$He($\alpha$+$n$).  Then the partial-wave
$\alpha$-$^2n$ interactions produce $^6$He($\alpha$+$^2n$), which can be
viewed as an effective interaction, since it accounts for omitted
intrinsic structure of the dineutron. 
The $^7$He($\alpha$+$n$+$^2n$) system also needs the $n$-$^2n$ partial wave
effective interaction. Then $^7$He could get contributions from more
than one intrinsic structure of $^2n$, where most is accounted for by
the effective interaction.  The $^8$He($\alpha$+$^2n$+$^2n$) system needs
the $^2n$-$^2n$ partial wave effective interaction, which also accounts for
the intrinsic $^2n$ structures, but not another unlikely cluster
divisions like $\alpha$+$n$+$^3n$.

The purpose of this work is to investigate how the three-body
structures, $\alpha$+$^2n$+$^2n$ and $\alpha$+$n$+$^2n$, permit to
describe the known properties of $^8$He and $^7$He, respectively.  We
treat the dineutron, $^2n$, as a zero-spin particle, with a typical
size, root-mean-square radius, of about 2 fm \cite{kan07}.  The main
uncertainty comes from the choices of $^2n$-$^2n$ and $n$-$^2n$
interactions.  For this reason particular emphasis will be put on what
the main characteristics of these interactions have to be such that
the $^8$He and $^7$He properties are fairly well reproduced.

\subsection{From data to predictions}

The experimental data on the weakly bound neutron dripline nuclei,
$^7$He and $^8$He, are binding energies, spins and parities of ground
and excited (resonance) states, and some of their widths.  In addition,
for $^8$He the root-mean-square radius is also known as well as the
missing mass spectrum after gentle removal of an $\alpha$-particle.
Two levels are known for $^7$He, that is $\frac{3}{2}^-$ and
$\frac{5}{2}^-$, and five levels for $^8$He, that is two $0^+$, and
one of each $2^+$, $1^-$, and $3^-$.

We shall use the three-body cluster model to describe $^7$He and
$^8$He structures.  The technique is the three-body hyperspherical
expansion method combined with complex scaling for resonance
calculations.  First the cluster divisions with the three constituents
are selected, that is $^7$He($\alpha$+$n$+$^2n$) and
$^8$He($\alpha$+$^2n$+$^2n$).  Then the input data are the two-body
interactions between the three subsystems, where each is a nucleus in
itself, that is $^5$He and $^6$He in addition to the nuclear
combinations $n$-$^2n$ and $^2n$-$^2n$. In this work
the spatial form of the two-body potentials are sums of Gaussian functions, usually
only one or two are sufficient.

The strength and range of the Gaussian potentials, together with the
angular momentum and spin
dependencies, are carefully chosen to reproduce the known properties of
the corresponding nuclei.  Both spatial and angular momentum
dependencies are chosen to be consistent with the Pauli exclusion
principle.  

To determine the proper interactions between unstable
constituents, $^2n$-$^2n$ and $n$-$^2n$, where direct two-body
information is not available, is more delicate. However, as constituents of a three-body
system, these interactions influence the resulting three-body
structure.  Information about three-body properties can then, through
the calculations, provide information about the two-body input
interaction.  Therefore, such interactions are intricately connected
with the three-body calculation.

The accuracy of the calculations are of few-body character, which
means that inaccuracies arise from choices of cluster division and how
well the two-body interactions can be determined from the available data.
The numerical accuracy is insignificant compared to the model choice.
The cluster constituents are crucial and chosen from inspection of
internal and external separation energies, which strongly must suggest
a small spatial overlap.  A wrong choice would be catastrophic and the
results would not be possible to repair by use of adjusted
phenomenological interactions.  The two-body interactions contain
uncertainties as well, but the two-body data combined with three-body
data and calculations are rather efficient in narrowing down their
uncertainties.  In this way, successful three-body calculations can
supply information about these unknown interactions.

The results of these cluster calculations must be in fair agreement
with much of the available data, otherwise the employed clusterization
is not preferred by nature. Good or bad agreement is significant to
assess how well the cluster division can describe the investigated
structures.  Predictions of so far unknown properties clearly also
provide possible tests.  This includes the detailed structures
obtained from the partial wave decomposition of bound state sand
resonances.  No matter how well the models turn out, they can never be
equally accurate for all properties of all states. In case the
energies are essential for example close to a threshold, we introduce
the three-body potential for fine tuning.  This does not change the
structure, only lowering or lifting the corresponding energy.
However, it is an indication of how well the model is doing for that
state.

The paper is organized as follows.  After the introduction and the
brief explanation of the overall procedure, we describe in Section II
the basic interactions related to the experimentally known subsystems
$^5$He($\alpha$+$n$) and $^6$He($\alpha$+$^2n$).  Section III
investigates various dineutron-dineutron interactions and their
effect, in comparison with measurements, on the three-body $0^+$
ground state of $^8$He($\alpha$+$^2n$+$^2n$).  The best interactions
are used in Section IV to compute excited states of
$^8$He($\alpha$+$^2n$+$^2n$).  In Section V, we turn to
$^7$He($\alpha$+$n$+$^2n$), where much less is known.  Finally,
Section VI summarizes and gives the conclusions.

\section{The $\alpha$-$^2n$ and $\alpha$-$n$ interactions}
The experimentally known subsystems, $^5$He and $^6$He, can first be
used to determine the two-body interactions.  These potentials are
assumed to be partial-wave dependent, and they contain a central, and
when needed, a spin-orbit part:
\begin{equation}
V^{(\ell)}(r)=V_c^{(\ell)}(r)+V_{so}^{(\ell)}(r) \bm{\ell}\cdot \bm{s},
\end{equation}
where $\ell$ is the relative orbital angular momentum between the two particles, and $s$ is the total spin.
The potential radial functions, $V(r)$, are considered to have a Gaussian shape, 
$V(r)=Se^{-r^2/b^2}$, whose strength, $S$, and range, $b$, are determined such that they reproduce the
available experimental information of the two-body system. 

\paragraph{The $\alpha$-dineutron potential:}  We here use the same $\alpha$-$^2n$ potential as the one
described in reference~\cite{gar25} and employed for the description of $^8$Li as an $\alpha$+$t$+$^2n$ structure.

\begin{table}[t!]
	\caption{Strength, $S$, and range, $b$,  in MeV and fm, respectively, for the Gaussian potentials used for the $\alpha$-$^2n$ and $\alpha$-$n$ interactions.}
\label{tab:potalpha}
\begin{indented}
\item \begin{tabular}{@{}l|ll|llll} \hline
  & \multicolumn{2}{c}{$\alpha$-$^2n$}  & \multicolumn{4}{|c}{$\alpha$-$n$} \\ 
  & $V_c^{(s)}$  & $V_c^{(d)}$&   $V_c^{(s)}$  & $V_c^{(p)}$  &   $V_c^{(d)}$ &  $V_{so} $\\ \hline
$S$  (MeV) & $-20.37$  &  $-63.35$ &  $48.0$ &  $-47.4$  &   $-21.93$  &  $-25.49$ \\
$b$ (fm) & $1.8$        &  $2.5$      &  2.33 &  2.30  &   $2.03$  &  1.72 \\  \hline
\end{tabular}
\end{indented}
\end{table}

This potential is based on the fact that the three-body structure of the $^6$He ground state corresponds to 
a very large extent with a relative $s$-wave between the two neutrons and a relative $s$-wave between the 
$\alpha$-particle and the dineutron center of mass \cite{gar99}.  This fact permits to obtain the parameters
of the  $s$-wave $\alpha$-$^2n$ central potential in such a way that the two-neutron separation energy is reproduced 
($-0.975$ MeV \cite{bro12}). The corresponding values for the strength and the range of the Gaussian $s$-wave
potential are given in the second column of Table~\ref{tab:potalpha}.

For the $d$-wave interaction we proceed in the same way, fitting the potential parameters in order to reproduce
the properties of the $2^+$ resonance in $^6$He. More precisely,  the strength and range of the $d$-wave
$\alpha$-$^2n$ potential are adjusted to reproduce the experimental resonance energy  and width, 0.83 MeV above the two-neutron threshold 
and 0.113 MeV, respectively \cite{til02}. The values of these parameters are given in the third column of Table~\ref{tab:potalpha}. It is important to note that the picture employed here assumes that the  $2^+$ resonance wave function is fully given by the component
having the two neutrons in a relative $s$-wave and the $\alpha$-particle in a relative $d$-wave with respect to the dineutron
center of mass. However,  three-body calculations show that the component with a relative $d$-wave between the two
neutrons and a relative $s$-wave between the $\alpha$-particle and the dineutron, although it gives a small contribution, its weight is
not negligible \cite{fed03}.  The effective interaction accounts for this assumption by reproducing the $2^+$ energy.

As discussed in reference~\cite{gar25}, a $p$-wave $\alpha$-$^2n$ interaction should in principle  be considered as well. However,  
the absence of a known low-lying $1^-$ state in $^6$He, and the fact the $p$-wave $\alpha$-$^2n$ components
play a very minor role in the three-body description of the   $0^+$ and $2^+$ states in $^6$He leads us to simply employ
for the $p$-wave potential the same interaction as for $s$-waves. In any case, as mentioned in \cite{gar25}, the
use of this $p$-wave potential, making it equal to zero, or even excluding the $p$-wave components from the
calculations, do not produce significant changes in the description of the low-lying $^8$Li states.

\paragraph{The $\alpha$-neutron potential:} For the $\alpha$-$n$ interaction we use the one specified in reference~\cite{cob98},
and whose parameters for the $s$- $p$- and $d$-wave central potential as well as the $\ell$-independent spin-orbit
potential are given in the right part of Table~\ref{tab:potalpha}.  These parameters have been obtained in order to
reproduce the $s$-, $p$-, and $d$-wave $\alpha$-neutron phase shifts obtained from scattering experiments up to 
an energy of 20 MeV. Furthermore, these parameters give rise to $\frac{3}{2}^-$ and $\frac{1}{2}^-$ resonances
whose energy and width agree with the experimental values \cite{til02,cob98}. Note that the $s$-wave potential is
repulsive, which guarantees that the neutron does not occupy the Pauli forbidden $s_{1/2}$-shell filled by the two 
neutrons in the $\alpha$-particle.

\section{The $^8$He ground state $0^+$}
\label{sec3}

Let us start by investigating the ground state, $0^+$, of $^8$He assuming an $\alpha$-$^2n$-$^2n$ structure.
The three-body calculations are made following the hyperspherical adiabatic expansion method described
in detail in \cite{nie01}.  They have been made including all the partial waves consistent with $\ell_x,\ell_y \leq 3$,
where $\ell_x$ and $\ell_y$ refer to the relative orbital angular momentum between any pair of particles and
$\ell_y$ is the orbital angular momentum between the third particle and the center of mass of the first two.
As mentioned above, and although the $p$- and $f$-waves play a minor role, 
for the $p$-wave interaction  between the $\alpha$ and the dineutron we use the same potential as
for $s$-waves, whereas for $\ell=3$ we use the same potential as for $d$-waves.

The three-body calculation provides the root mean square (rms) radius given by % \cite{?}:
\begin{equation}
\langle r^2 \rangle =\frac{\langle \rho^2 \rangle}{A}+\sum_{i=1}^3\frac{A_i}{A}\langle r_{ci}^2 \rangle,
\label{rms}
\end{equation}
where $A_i$ is the mass number of cluster $i$, $A=\sum_{i=1}^3A_i$ is
the mass number of the full system, $\langle r_{ci}^2 \rangle^{1/2}$
is the rms radius of cluster $i$, and $\langle \rho^2 \rangle$ is the
expectation value of the square of the hyperradius $\rho$, defined in
\cite{nie01} with normalization mass as that of the nucleon.

\subsection{The $^2n$-$^2n$ interaction}

In order to perform the calculations, the $^2n$-$^2n$ interaction still has to be specified. In reference~\cite{hag08} the
dineutron correlation in the ground state of $^8$He was investigated using a core+$4n$ model. They found that 
the neutron pairs coupled to zero spin exhibit a strong dineutron correlation, and furthermore, they
also concluded that the two dineutrons interact very weakly with each other. For this reason we start 
our three-body calculation simply putting the $^2n$-$^2n$ interaction equal to zero.

\begin{table}[t!]
\caption{Strength, $S$, and range, $b$, of the four $^2n$-$^2n$ potentials used in this work. For each of them, and for the $0^+$ ground state of $^8$He, we also give the three-body binding energy, $B$, and rms radius, $\langle r^2\rangle^\frac{1}{2}$,
when no three-body potential is used,  the strength, $S_\mathrm{3b}$, of the three-body potential ($b_\mathrm{3b}=4.5$ fm) needed to fit the experimental binding
energy ($-3.11$ MeV \cite{til04}), and the rms radius, $\langle r^2\rangle^\frac{1}{2}_\mathrm{3b}$, obtained when the three-body potential is included. All the energies
are given in MeV and the distances in fm.}
\label{tab:gs}
\begin{indented}
\item
\begin{tabular}{@{}l|llll|llll} \hline
  & \multicolumn{2}{c}{$s$-wave}  &  \multicolumn{2}{c|}{$d$-wave} & & &  &  \\
  & $S$   &   $b$  & $S$   &   $b$   & $B$  &  $\langle r^2\rangle^\frac{1}{2}$ & $S_\mathrm{3b}$  &  $\langle r^2\rangle^\frac{1}{2}_\mathrm{3b}$ \\ \hline
Potential I  & 0  &  $--$&  0 &  $--$   & $-2.20$ &  2.80 &  $-2.48$ &  2.58  \\
Potential II & $-8.3$  &  1.5 &  $-8.3$ &  1.5 &$-3.11$ & 2.54 & 0 &  2.54  \\   
Potential III& $30.0$  &  1.5 &  $-220.0$ &  1.5 & $-2.07$ & 2.75  & $-3.25$  & 2.61  \\
Potential IV& $30.0$  &  1.5 &  $-229.8$ &  1.5 & $-3.11$ &   & 0 &  2.57 \\ \hline
\end{tabular}
\end{indented}
\end{table}

When this is done, we obtain a bound $0^+$ state, with a four-neutron separation energy of 2.20 MeV, about 0.9 MeV
smaller than the experimental value of 3.11 MeV \cite{til04}. A not very much attractive three-body force is sufficient
to fit the computed energy to the experimental value. In particular, if the Gaussian three-body force
\begin{equation}
V_\mathrm{3b}=S_\mathrm{3b} e^{-\rho^2/b_\mathrm{3b}^2}
\label{gau3b}
\end{equation}
is used, and we choose a range $b_\mathrm{3b}$=4.5 fm, a strength $S_\mathrm{3b}=-2.48$ MeV permits then to recover the 
experimental energy. 

The root mean square radius is also a measurable quantity. 
In our case, since $\langle r^2_\alpha \rangle^{1/2}=1.67$ fm \cite{kra20}, and considering 
$\langle r^2_{2n}\rangle^{1/2}\approx 2$ fm \cite{kan07}, we obtain from equation~(\ref{rms}) a rms radius for the $^8$He ground
state of 2.80 fm and 2.58 fm  without and with inclusion of the three-body force, respectively.
When compared to the experimental value, $\langle r^2\rangle ^{1/2}=2.53\pm 0.02$ fm \cite{wak22},
we see that the inclusion of the three-body force, besides the four-neutron separation energy, gives rise to a 
rms radius also in reasonable good agreement with the experiment.
The results obtained with this potential, referred as potential I, are collected in the first line of Table~\ref{tab:gs}.

The fact that a zero interaction between the two dineutrons gives rise to a separation energy of the $^8$He 
ground state not dramatically different from the experimental value, leads us to consider the case where the
$^2n$-$^2n$ interaction is slightly attractive, simply to reduce the role of the three-body force. Playing with the strength
and the range of the Gaussian potential, it is possible to get  values that permit to reproduce simultaneously the experimental 
separation energy and rms radius without need of any three-body force. This happens 
with the strength and range $S=-8.3$ MeV and $b=1.5$ fm. This potential will be referred as potential II, and the 
computed results for the $0^+$ ground state are given in the second row of Table~\ref{tab:gs}. 

In reference~\cite{ber03} the authors investigated the properties of the tetraneutron as a dineutron-dineutron molecule.
In their work they estimated that the $s$-wave dineutron-dineutron potential is predominantly repulsive. For this reason 
we have considered as well the possibility of a repulsive $s$-wave $^2n$-$^2n$ potential and an attractive $d$-wave potential
that permits to recover the correct three-body separation energy. These are the potentials denoted as 
III and IV in Table~\ref{tab:gs}. They have been constructed similar to potentials I and II, respectively. For potential III
some three-body attraction is needed to fit the experimental separation energy, and in potential IV the $d$-wave
interaction is used to adjust the energy and the three-body force is not needed. As we can see, these potentials, III and
IV, produce a slightly bigger ground state, although still in reasonable agreement with the experimental value.  

In the very recent paper \cite{nak25}, the authors have investigated the
structure of the two known $0^+$ states in $^8$He.  They allowed both
dineutrons and shell-model $p_{3/2}$-configurations.  The separation
between two-body, three-body and shell-model structure is defined
through the choice of a non-orthogonal, and substantially
overlapping, basis. They find $\alpha$+$^2n$+$^2n$ configurations (their
two- and three-body clusters) in the $0^+$ ground state with probability $0.883$. Their
shell model consists of two neutrons in $p_{3/2}$-configurations, which
to some degree could also be describing a dineutron.  This result supports the
$\alpha$+$^2n$+$^2n$ description used here. Furthermore, in \cite{nak25} they 
obtain a rms radius of 2.49~fm, in fairly good agreement with our result of about
2.6~fm obtained with the four potentials, as shown in the last column of 
Table~\ref{tab:gs}, and with the experimental value of $2.53\pm
0.02$~fm. In any case, the large amount of $\alpha$+$^2n$+$^2n$ configuration
found in \cite{nak25} contrasts with the result in reference~\cite{yam23}, where a rather small 
squared overlap of about 45\% with
the $\alpha$+$^2n$+$^2n$ configuration is assigned to the $0^+$ ground state.

\begin{table}[t!]
\caption{For the $0^+_1$ state in $^8$He, and for each of the  two different Jacobi sets, contribution to the norm of the wave function of the components 
included in the calculation.
The results with the four $^2n$-$^2n$ potentials in Table~\ref{tab:gs} are given. When needed the three-body potential (\ref{gau3b}) has been 
included in the calculation.}
\label{tab:gscomp}
\begin{indented}
\item
\begin{tabular}{@{}lllllll|lllllll}\hline
\multicolumn{14}{c}{0$^+_1$ state (ground state)}  \\ 
\multicolumn{7}{c}{$4n$($^2n$,$^2n$)+$\alpha$}  & \multicolumn{7}{c}{$^6$He($\alpha$,$^2n$)+$^2n$}  \\\hline
 $\ell_x$   & $\ell_y$ &$L$  &  \multicolumn{4}{c|}{weights} &  $\ell_x$ &$\ell_y$   &  $L$ &  \multicolumn{4}{c}{weights} \\ \cline{4-7} \cline{11-14}
   &    &    & I &  II & III &  IV &  &   &    &  I & II&  III  & IV \\  
     0    &  0   &    0  &  96.9  &  96.7 & 44.3& 35.3 & 0  & 0  & 0 &  96.2 & 96.6 & 37.1& 33.8\\
     2    &  2   &    0  & 3.1 & 3.3 & 55.7&  64.7& 2  & 2  & 0 & 1.4 & 2.2 & 51.7& 53.1\\
           &      &       &    &   &   & & 1  & 1  & 0 & 2.3 & 1.2 & 9.6  & 10.7 \\
           &      &       &    &   &   & & 3  & 3  & 0 &       &       &  1.6  & 2.3 \\
\hline
\end{tabular}
\end{indented}
\end{table}

\subsection{Evaluating the potentials}

We have then constructed a set of four different $^2n$-$^2n$ potentials all of them reproducing the experimental ground state
three-body separation energy in $^8$He (using an attractive three-body force when needed), and also reproducing reasonably well the experimental
rms radius.  However, the structure of the state produced by these potentials is very different. This is shown in Table~\ref{tab:gscomp}, where
we give the contribution to the norm of the ground state wave function of the components  included in the calculation. 
The quantum numbers $\ell_x$ and $\ell_y$, which couple to $L$, refer to the relative orbital angular momenta between two of the particles, and between
the third particle and the center of mass of the first two, respectively. The results in the two possible Jacobi sets, with the $\bm{x}$-Jacobi coordinate  between
 the two dineutrons or between the $\alpha$-particle and one of the dineutrons, are given in the left and right parts of the table. As we can see, potentials 
 I and II favor the contribution of the $s$-wave components in both Jacobi sets, which give about 97\% of the wave function.  Thus, this $^8$He state is mostly the ground state of $^6$He$(0^+)$ supplemented by another $s$-wave dineutron, but the structure can as well in the other basis be described as an alpha-particle surrounded by two dineutrons in a relative s-wave.

On the contrary, potentials III and IV enhance significantly the role
of the $d$-waves, which in both Jacobi sets contribute with at least
50\% of the norm. Thus, now the $^8$He structure is equally shared
between $^6$He$(0^+)$ and excited $^6$He*$(2^+)$ states again
supplemented by another dineutron.  Furthermore, in these cases, the
$p$-wave component between the $\alpha$-particle and the dineutron has
a relevant contribution of about 10\%.  The similarity between the
results for potentials I and II, and for potentials III and IV,
indicate that the need or not of the three-body force is not playing
an important role, at least with respect to the partial wave content in
the wave function.

\begin{figure}[t!]
\centering
\includegraphics[width=12cm]{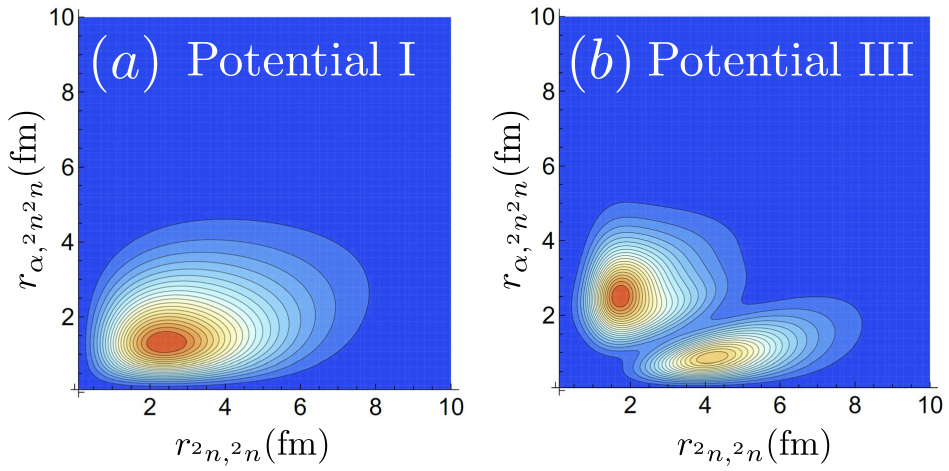}
\caption{Contour plot of the $0^+$ ground state with $^2n$-$^2n$ potentials I and III as a function of the $^2n$-$^2n$ distance and the distance between the 
$\alpha$-particle and the center of mass of the two dineutrons. }
\label{fig:cont}
\end{figure}

The different structure produced by the potentials can also be seen in
figure~\ref{fig:cont}, where we show the contour plot of probability
function as a function of the $^2n$-$^2n$ distance and the distance
between the $\alpha$-particle and the center of mass of the two
dineutrons. Only the results with potentials I and III are shown,
since the plots with potential II and IV are indistinguishable by
naked eye from the ones with potentials I and III, respectively. The
probability distribution with potential I (and II) is elongated along
the dineutron-dineutron distance with a single maximum with the
$\alpha$-particle about 1~fm away from the $^2n$-$^2n$ center of
mass. This is a relatively tight structure, although the three
particles still remain separated.  However, for potential III (and IV)
the distribution presents two peaks, one of them also elongated along
the $^2n$-$^2n$ distance, but the second one, which actually
dominates, appears for clearly closer dineutrons to each other and the
$ \alpha$-particle further away from the two-dineutron center of mass.

\begin{figure}[t!]
\centering
\includegraphics[width=10cm]{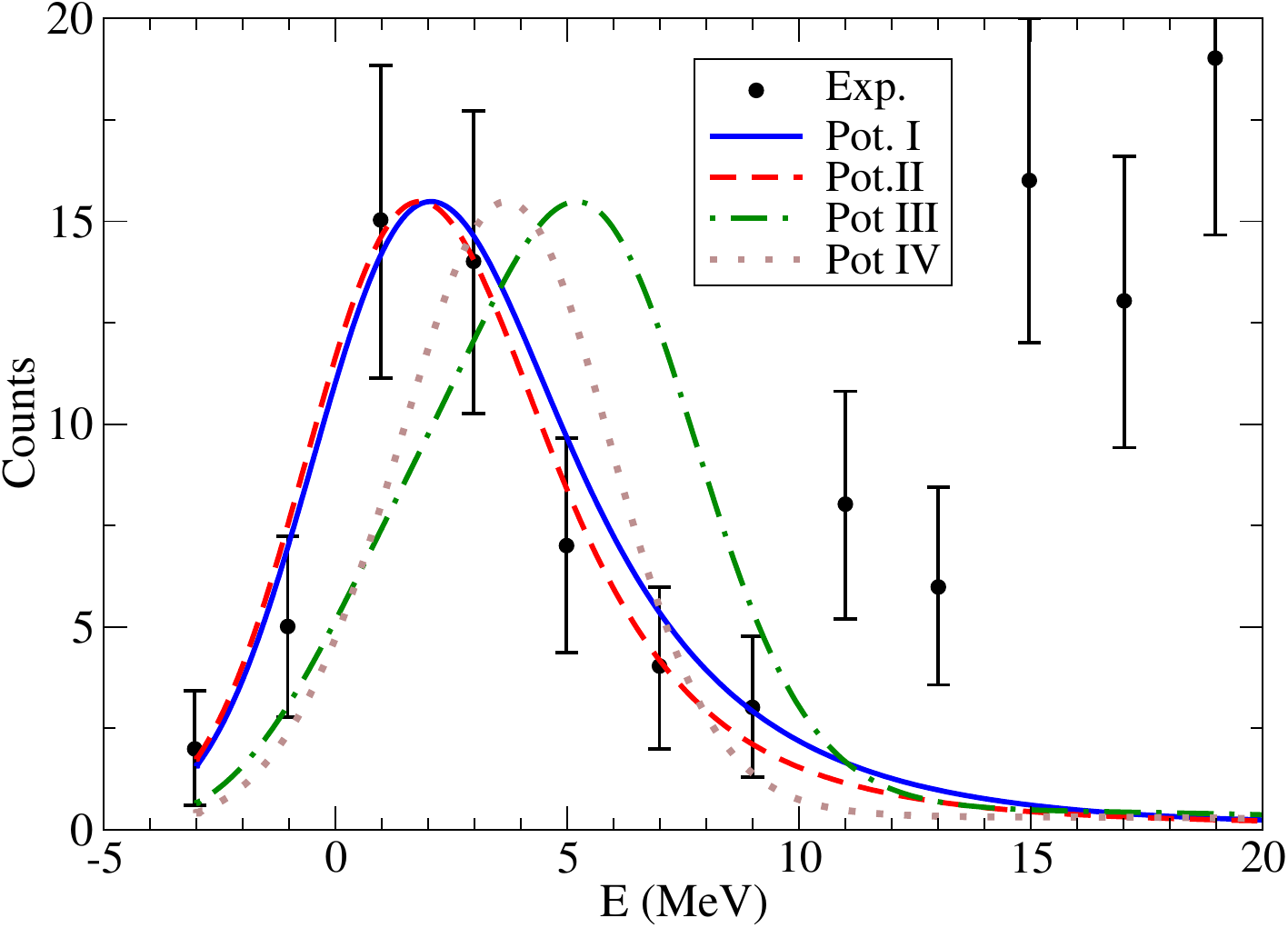}
\caption{Missing-mass spectrum  after removal of the $\alpha$-particle in the $\alpha$+$^2n$+$^2n$ system  obtained with the four $^2n$-$^2n$ potentials 
given in Table~\ref{tab:gs}. Experimental data from \cite{due22}.}
\label{fig1}
\end{figure}

\subsection{Missing-mass spectrum}

Very recently, the missing-mass spectrum of the four-neutron system after the 
$^8$He($p$,$p\alpha$)$4n$ reaction has been measured at a $^8$He collision energy of 156 MeV/nucleon \cite{due22}. 
As a test, we can estimate this spectrum
by means of the sudden approximation, which assumes that in the collision one of the three-constituents
of the system (the $\alpha$-particle in our case) is suddenly removed, is such a way that the other two particles
(the two dineutrons) continue flying together without being too much altered by the removal of the core.
As shown in \cite{gar97}, the momentum distributions of the fragments is very much determined by the final 
state interaction between them. Taking this into account, and following \cite{gar97b}, we have computed
the invariant mass spectrum for $^8$He after removal of the core. The result is shown in figure~\ref{fig1}
for the four $^2n$-$^2n$ potentials considered in this work. The computed curves have been convoluted with an
experimental resolution of 2 MeV (which is assumed to be energy independent) \cite{laz23} and scaled to the same
maximum. 

As we can see, potentials I and II both give rise to spectra reproducing reasonably well 
the low-energy peak of the experimental spectrum. On the contrary, the spectra produced by potentials III and IV show a peak 
clearly shifted with respect to the experimental data.  

The fact that the spectrum with potential IV (for which there is no need 
of including a three-body force in the three-body calculation) gets closer to the experimental peak than with potential III
(for which a three-body attractive potential is needed) could lead us to think that a more attractive $d$-wave $^2n$-$^2n$ potential
and a repulsive three-body force would reproduce better the experimental spectrum in figure~\ref{fig1}.
However, there is not much room for this, since a $d$-wave attractive
strength of about $-240$ MeV binds the two dineutrons into an
unphysical $2^+$ four-neutron bound state. Therefore, from this
perspective, the dineutron-dineutron potentials showing a very weak
interaction are more appropriate than the ones containing
repulsive $s$-wave and attractive $d$-wave terms. 

In any case, experimentally the energy $E$ in the
figure refers to the four-neutron relative energy, whereas in the
calculations $E$ represents the dineutron-dineutron relative energy.
The four-neutron relative energy equals the dineutron-dineutron
relative energy added to twice the intrinsic $n$-$n$ energy, which
in turn should be slightly positive.  A guess could be the $n$-$n$
virtual energy of $\hbar^2/(m a_{sc}^2)\simeq 0.12$~MeV
(where the $n$-$n$ scattering length $a_{sc}=18.7$ fm has been used).
The curves
should then be shifted by about 0.24 MeV to the left in figure~\ref{fig1},
which would be hardly noticeable.

\section{$^8$He excited states}

Experimentally, all the excited states of $^8$He are known to be resonances \cite{til04}. In order to get them
numerically, the calculations are made after complex rotation of the hyperspherical adiabatic
expansion method mentioned above, as done for instance in \cite{gar02}. In this way, the resonances appear 
as states with bound state asymptotic behavior, but with complex energy, $E=E_R-i\Gamma_R/2$, where 
$E_R$ and $\Gamma_R$ are the resonance energy and width, respectively.

\subsection{$0^+$ states}

\begin{figure}[t!]
\begin{center}
\includegraphics[width=9cm]{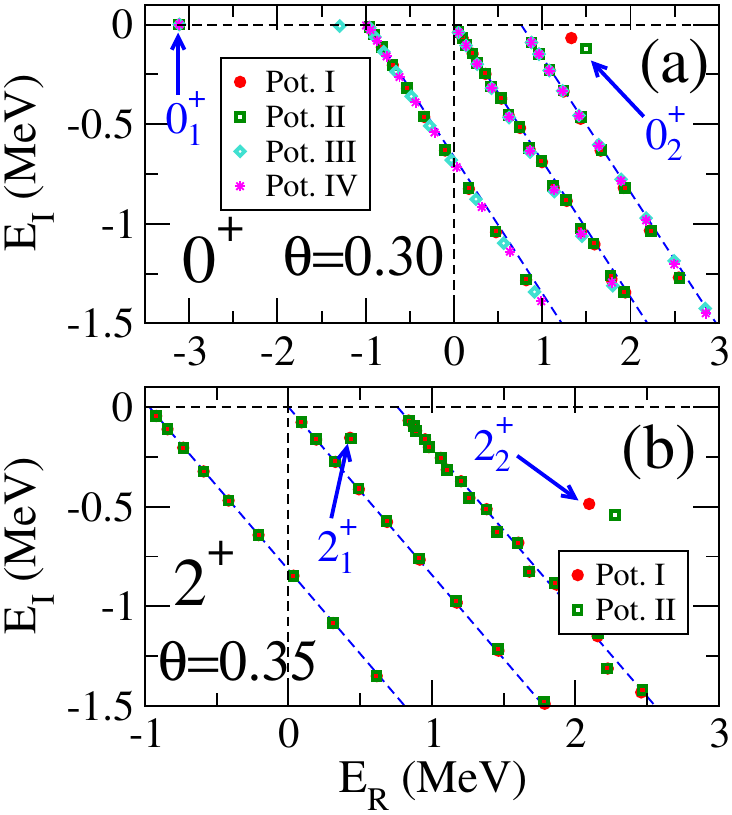}
\caption{Complex scaled energy spectrum of the (a) $0^+$ (with complex rotation angle $\theta$=0.30 rads)  and (b) $2^+$  states  (with $\theta$=0.35 rads) 
obtained with the potentials given  in Table~\ref{tab:gs}. }
\label{fig2}
\end{center}
\end{figure}

As mentioned above, the possibility of existence of excited $0^+$ states is investigated by solving the three-body problem after a complex 
scaling transformation \cite{gar02}. More precisely, this is done after discretization of the continuum spectrum by imposing a box boundary condition. 

The result for the $0^+$ states is shown in figure~\ref{fig2}a, where a complex scaling angle $\theta$=0.30 rads has been used.
The discrete three-body energies shown in the figure are distributed along three
different cuts, which are rotated by an angle $2\theta$. The first cut starts at the $^6$He two-neutron separation energy ($-0.97$ MeV), and the dots
along this cut 
represent $^6$He-$^2n$ continuum states. The second cut, which starts at zero energy, represents genuine
three-body continuum states. Finally, the third cut starts at 0.82 MeV, which is the energy of the $2^+$
resonance in $^6$He, and the dots correspond to a sort of two-body continuum states between the $2^+$ resonance
in $^6$He and the dineutron. The dots out of these three cuts, indicated by the arrows in the figure, are the
bound and resonant $0^+$ states in $^8$He. For each of the potentials the calculation has been made using the same three-body force as
the one needed to  reproduce the energy of the ground state. 

\begin{table}[t!]
\caption{Energies, $E_R$, and widths, $\Gamma_R$, of the computed $^8$He states with the four $^2n$-$^2n$
potentials given in Table~\ref{tab:gs}. The energies (with respect to the  three-body threshold) and the widths are given in MeV.  For 
each state we also give the strength, in MeV, of the three-body force (\ref{gau3b}) used in the calculation.
The experimental data have been taken
from \cite{til04} except the one of the $0^+_2$ state, which has been taken from  \cite{yan23}.}
\label{tab2}
	\setlength\tabcolsep{.15\tabcolsep}
\begin{indented}
\item
\begin{tabular}{@{}c|ccccccccccccccc}\hline
  &  \multicolumn{3}{c}{Pot. I} & \multicolumn{3}{c}{Pot. II} & \multicolumn{3}{c}{Pot. III} &  \multicolumn{3}{c}{Pot. IV} & \multicolumn{2}{c}{Exp.}  \\ \hline
  & $E_R$   &  $\Gamma_R$  & $S_\mathrm{3b}$ & $E_R$   &  $\Gamma_R$  &  $S_\mathrm{3b}$ &   $E_R$   &  $\Gamma_R$ &    $S_\mathrm{3b}$ &  $E_R$   &  $\Gamma_R$ &$S_\mathrm{3b}$ &$E_R$   &  $\Gamma_R$     \\
 $0^+_1$   &   $-3.11$   &   ---       &$-2.48$  & $-3.11$   & ---      & 0.0 & $-3.11$  &  ---  & $-3.25$& $-3.11$ & ---& 0.0 &$-3.11$  &  ---  \\
 $0^+_2$   &   $ 1.33$   &   0.13   &$-2.48$ &  1.50        & 0.24   & 0.0 & $-1.29$  &  --- & $-3.25$& $-1.0$    & ---  &    0.0       &1.43$\pm$0.06 & \\
  $2^+_1$   &   $0.43$   &   0.31   & 5.0 & $0.43$   & 0.31   & 7.25 &                  &        &  &                &       &                & $0.43\pm 0.06$&   0.89$\pm$0.11\\ 
  $2^+_2$   &   $ 2.10$  &   0.97   & 5.0 &  2.27       &  1.09  & 7.25&                   &       &   &              &         &                &  &  \\
  $1^-_1$   & 0.33          & 1.35      & $-6.0$&  0.19       & 1.39   & $-6.0$ &                   &      &    &              &         &                &   1.25$\pm$0.20 &  1.3$\pm$0.5 \\         
  $3^-_1$   & 0.70          & 0.23     & 0.0&  0.01        & 0.23   & 0.0 &                  &       &    &             &           &               &   4.05$\pm$0.04 & 0.1$\pm$0.1 \\       \hline
 \end{tabular}
\end{indented}
\end{table}

As we can see, together with the bound state at $-3.11$ MeV, potentials I and II show a second, resonant, $0^+$ state with $E_R$=1.33 MeV and
1.50 MeV above the four-neutron threshold  (therefore with excitation energies 4.44 MeV and 3.61 MeV, respectively). The computed width of the resonance
is 0.13 MeV and 0.24 MeV for each of the two potentials. This resonance excitation energy  agrees very well with the one recently observed in reference~\cite{yan23}, where they have
identified a $0^+$ resonance in $^8$He with excitation energy 4.54$\pm$0.06 MeV, which corresponds to an
energy of 1.43 MeV above the four-neutron threshold. In \cite{yan23} this state is interpreted as an exotic $\alpha$+$^2n$+$^2n$ cluster structure. 
Nothing is said in \cite{yan23} concerning the width of the $0^+_2$ state. This $0^+$ excited state has also been obtained in \cite{kob13,myo10} with
excitation energies of 4.5 MeV and 6.3 MeV, respectively. In \cite{myo10} a width of 3.2 MeV is given, clearly 
bigger than the values provided by potentials I and II.

However, when potentials III and IV are considered, we have found that the $0^+_2$ state is actually bound, as  clearly seen in figure~\ref{fig2}a for potential III, where
a second bound state is observed at $-1.29$ MeV below the three-body threshold. For potential IV the energy of the $0^+_2$ state is $-1.0$ MeV, so close to the
$^6$He($0^+$)+$^2n$ threshold that the state can not be seen in the figure. In any case, these two energies clearly disagree with the experimental value of
1.43 MeV. Even more, if we try to force the energy of the $0^+_2$ state to reproduce the experimental energy by means of a sufficiently repulsive three-body force,
we then find that the $0^+_2$ state disappears very fast in the $^6$He($0^+$)+$^2n$ continuum, and the experimental energy can not be reached.  

\begin{figure}[t!]
\centering
\includegraphics[width=7cm]{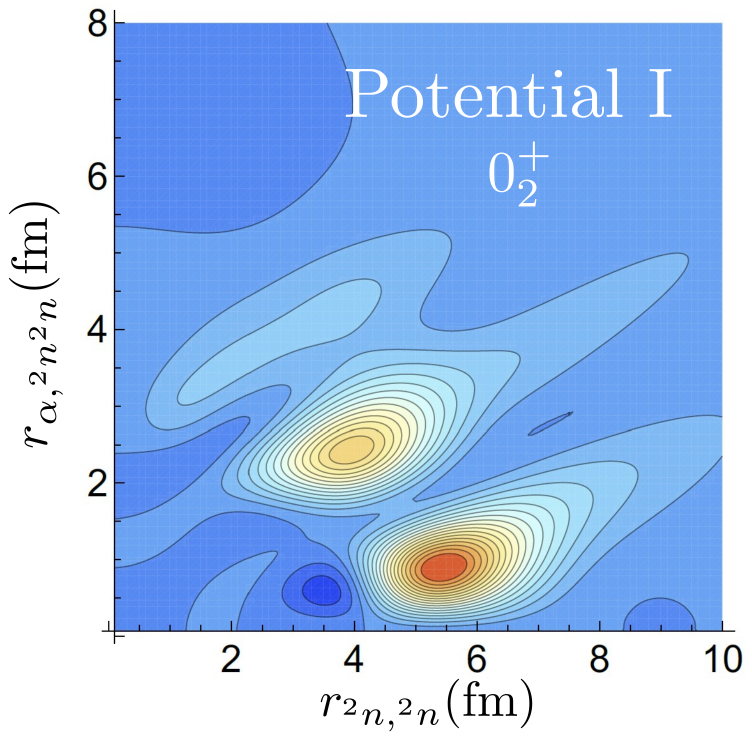}
\caption{The same as figure~\ref{fig:cont}  for the excited $0^+$ resonance.}
\label{fig:cont2}
\end{figure}

In the first two rows of Table~\ref{tab2} we collect the computed energies and widths for the $0^+_1$ and $0^+_2$ states using the same three-body force as
the one needed to reproduce the ground state energy. It is clear that potentials I and II
are more appropriate for a correct description of the $0^+_2$ state, since a small retuning of the three-body force would be sufficient to recover
the experimental energy.

Although the resonances are obtained after a complex scaling calculation, the real part of the squared complex rotated wave function can give a clear hint about the spatial
distribution of the clusters when the resonance is populated. The contour plot of this probability function is shown in 
figure~\ref{fig:cont2} for the $0^+_2$ resonance with potential I (the result obtained with potential II is very similar). As in figure~\ref{fig:cont}, we plot it as a function
of the $^2n$-$^2n$ distance and the distance between the  $\alpha$-particle and the center of mass of the two dineutrons.
As we can see, the function presents two prominent peaks, where the main one is located for a large separation between the two dineutrons, about 5.5 fm, and the $\alpha$-particle
pretty close, about 1 fm, to the two-dineutron center of mass. When comparing to figure~\ref{fig:cont}a we can see that, as expected, the $0^+_2$ wave function is more
extended than the one of the ground state. This is also reflected in the rms radius obtained for the $0^+_2$ resonance, which is 3.1$\pm$0.6 fm with both, potentials I and II.
As in the case of the complex energy, the imaginary part of the computed rms radius is interpreted as the uncertainty of the observable.

\begin{table}[t!]
\caption{The same as Table~\ref{tab:gscomp} for the $0^+_2$ state.}
\label{tab:0+2}
\begin{indented}
\item
\begin{tabular}{@{}lllllll|lllllll} \hline
\multicolumn{14}{c}{0$^+_2$ state}  \\
\multicolumn{7}{c}{$4n$($^2n$,$^2n$)+$\alpha$}  & \multicolumn{7}{c}{$^6$He($\alpha$,$^2n$)+$^2n$}  \\\hline
 $\ell_x$   & $\ell_y$ &$L$  &  \multicolumn{4}{c|}{weights} &  $\ell_x$ &$\ell_y$   &  $L$ &  \multicolumn{4}{c}{weights} \\ \cline{4-7} \cline{11-14}
   &    &    &  I &  II &  III &  IV &   &   &    &   I &  II&   III & IV \\ 
     0    &  0   &    0  & 77.3    & 78.5  & 84.6& 65.3 & 0  & 0  & 0 & 10.9  & 11.2 & 79.7 & 75.4\\
     2    &  2   &    0  & 22.7  & 21.5 & 15.4  &  34.7 &  2  & 2  & 0 & 72.7 & 73.3 & 10.8 & 8.1 \\
          &        &       &       &      &        &   &   1  & 1  & 0 & 9.3 & 9.0 & 8.9 & 13.3\\
          &        &       &       &      &        &   &   3  & 3  & 0 & 7.1 & 6.5 & 0.6 & 3.2\\           \hline
\end{tabular}
\end{indented}
\end{table}

In Table~\ref{tab:0+2} we give the contribution to the norm of the different components in the $0^+_2$ state for the energies given in Table~\ref{tab2}.
We see that for potentials I and II the waves with $\ell>0$ start playing now a more  important role than for the ground state. In fact, from the right part of the table 
we can see that a large  contribution is coming from $^6$He in the $2^+$ state with the second dineutron in a relative $d$-wave. For potentials III and IV, which
fail in reproducing the experimental energy, it is just the opposite, since the dominating component corresponds to $^6$He in the ground state ($0^+$), and
the dineutron in a relative $s$-wave. We can also see that, for all the cases, the components
with odd value of $\ell$ give a significant contribution, which means that a more detailed analysis of the $p$- and $f$-wave $\alpha$-dineutron
potential than the one made here is very likely necessary to be more accurate about the structure of this state. 

The $0^+_2$ state in $^8$He was also investigated in
references~\cite{kan07,kob13,yam23} with the conclusion that there is a large
component of $\alpha$+$^2n$+$^2n$ structure.  In \cite{nak25} they
obtain an overlap probability of $0.926$ for this structure. They also
give a computed radius of $4.67$~fm for this first excited $0^+$ state in
$^8$He. As already mentioned, in our case this state is a resonance obtained by use of
complex scaling technique, where the imaginary part of the radius is
interpreted as an uncertainty resulting in the value $3.1 \pm 0.6$~fm.

The discretization of the continuum $0^+$ states allows us to compute the monopole transition strength to the ground state as:
\begin{equation}
%\frac{dB}{dE}=\sum_{k} \delta(E-E_k) |\langle \Psi_\mathrm{g.s.} ||\hat{\cal O} ||  \Psi_{0^+}(E_k)\rangle|^2,
{\cal B}=\sum_{k}  |\langle \Psi_\mathrm{g.s.} ||\hat{\cal O} ||  \Psi_{0^+}(E_k)\rangle|^2,
\label{monop}
\end{equation}
where $\Psi_\mathrm{g.s.}$ is the wave function of the $0^+_1$ ground state, $\Psi_{0^+}(E_k)$ is the wave function of the discretized continuum state 
$k$ with energy $E_k$, and $\hat{\cal O}=\sum_{i=1}^A r_i^2$ is the monopole transition operator \cite{kan16} (where $r_i$ is the radial coordinate of nucleon $i$ in the
nucleus, referred to the nucleus center of mass). 

The strength in equation~(\ref{monop}) can be computed on the real energy axis (without complex scaling). When the summation over the discrete states $k$ is limited to the 
ones whose energy is in the interval $E_R\pm 1$ MeV (where $E_R$ is the energy of $0^+_2$ resonance) we get ${\cal B}=113.4$ fm$^4$.  The square root of this value,  10.7 fm$^2$, 
compares well with the experimental value, $11^{+1.8}_{-2.3}$ fm$^2$, given in reference~\cite{yan23} for the monopole transition matrix element.

\subsection{$2^+$ states} 
Concerning the $2^+$ states, in reference~\cite{kor93} a $2^+$ resonance in $^8$He was identified with excitation energy
$3.57\pm 0.12$ MeV (0.46 MeV above the four-neutron threshold) and a width of $0.50 \pm 0.35$ MeV. These values have been 
recently confirmed in \cite{hol21}, where they have observed the resonance with excitation energy $3.54\pm 0.06$ MeV
(0.43 MeV above the four-neutron threshold) and a width of $0.89 \pm 0.11$ MeV. 

After the calculations,  and without the inclusion of any three-body potential, we have obtained that the $^2n$-$^2n$ potentials I and II give rise to a 
2$^+$ state at about 0.8 MeV  and 1.7 MeV, respectively, below the four-neutron threshold. Therefore, some repulsive three-body force is necessary
to push up the energy to the experimental value.  Using again the Gaussian three-body potential in equation~(\ref{gau3b}) 
with $b=4.5$ fm, this is achieved with the repulsive strengths $S_\mathrm{3b}$=5.0 MeV  and  $S_\mathrm{3b}$=7.25 MeV, for each of the 
two potentials. The corresponding discretized continuum spectra are shown in figure~\ref{fig2}b, which has been computed with the complex scaling angle $\theta=0.35$ rads.
In the figure we can see that both potentials give rise to a $2^+_1$ state with a very similar width, $\Gamma_R$=0.31 MeV,
which is close to a factor of 2 smaller than the experimental observation reported in  \cite{hol21}.

Concerning potentials III and IV, the results for the $2^+$ state are very different. In fact, they produce a lowest $2^+_1$ state, which actually would be the ground state, 
since it appears with a rather big three-body separation energy of about 20 MeV. Even more, they also show a pretty much  bound $2^+_2$ excited state, about 8 MeV below the 
four-neutron threshold. These results 
are so unrealistic, and so far from the known experimental information about $^8$He, that we definitely can conclude that
potentials III and IV must be disregarded as potentials suitable for the description of $^8$He as an $\alpha$+$^2n$+$^2n$ system. This is in fact 
confirming what we could already suspect, from the result shown in figure~\ref{fig1} and the fact that  the experimentally known $0^+_2$ state is not predicted
either when these two potentials are used.

In figure~\ref{fig2}b we can also see that, together with the $2^+_1$ state, a second $2^+$ resonance is found at about 2.2 MeV above the three-body threshold
(about 5.3 MeV excitation energy)  with a width in the vicinity of 1 MeV. So far there is no experimental evidence for the existence of such an excited 2$^+$ state.
In reference~\cite{myo21} a second $2^+$ resonance is also obtained, but at a higher energy (about 4.5 MeV above the four-neutron threshold) and clearly
wider (about 4.4 MeV). The precise computed energies and widths of the $2^+$ states obtained with potentials I and II are also given in Table~\ref{tab2}.

\begin{table}[t!]
\caption{The same as Table~\ref{tab:gscomp} for the $2^+_1$ and $2^+_2$ states with potentials I and II. }
\label{tab2+}
\begin{indented}
\item
\begin{tabular}{@{}lllllll|lllllll} \hline
\multicolumn{14}{c}{2$^+_1$ and $2^+_2$ states}  \\
\multicolumn{7}{c}{$4n$($^2n$,$^2n$)+$\alpha$}  & \multicolumn{7}{c}{$^6$He($\alpha$,$^2n$)+$^2n$}  \\\hline
 $\ell_x$   & $\ell_y$ &$L$  &  \multicolumn{2}{c}{$2^+_1$} & \multicolumn{2}{c|}{$2^+_2$}  & $\ell_x$ &$\ell_y$   &  $L$ &  \multicolumn{2}{c}{$2^+_1$} & \multicolumn{2}{c}{$2^+_2$}
 \\  \cline{4-5}  \cline{6-7} \cline{11-14}
   &    &    &  I &  II &  I &  II &   &   &    &   I &  II&   I & II \\ 
     0    &  2   &    2  & 63.7    & 67.9  & 24.0& 25.9 & 0  & 2  & 2 & 35.7  & 36.1 & 3.7 & 3.1\\
     2    &  0   &    2  & 34.1  & 31.8 & 27.7  &  22.6 &  2  & 0  & 2 & 54.8 & 56.9 & 3.7 & 3.4 \\
     2    &  2   &   2    & 2.3     &  0.3     &    48.3    &  51.4 &   2  & 2   & 2 & 2.6  &     0.6      & 81.0 & 83.1\\
          &        &       &       &      &        &   &   1  & 1  & 2 &  0.0     & 1.1 & 3.3 &   2.8  \\ 
          &        &       &       &      &        &   &   1  & 3  & 2 & 4.4 & 3.6 &  0.4     &    0.0  \\        
          &        &       &       &      &        &   &   3  & 1  & 2 & 2.0 & 1.4 &  0.2      &  0.4   \\      
          &        &       &       &      &        &   &   3  & 3  & 2 &  0.5  &    0.3    & 7.7  &  7.2   \\           \hline
\end{tabular}
\end{indented}
\end{table}

In Table~\ref{tab2+} we give the contribution of the different components in the two different Jacobi sets for the $2^+_1$ and
$2^+_2$ states with potentials I and II.  We see that the $2^+_1$ state has basically an $sd$ structure, with two of the particles in a relative $s$ (or $d$) wave, and the third particle in a  $d$ (or $s$) wave relative to the center of mass of the first two. In the first Jacobi set the relative $s$-wave between the two dineutrons dominate, giving about two thirds of the 
norm of the wave function. In the second Jacobi set the dominating components corresponds to $^6$He in a $2^+$ state.  For the $2^+_2$ state we find that the $\ell_x$=$\ell_y$=2
components are the dominant ones, mainly in the second Jacobi set, where they give more than 80\% of the norm. 
The difference between potentials I and II is not significant.

\subsection{$1^-$ and $3^-$ states}
The experimental information about $^8$He available in reference~\cite{til04} indicates that, although not fully confirmed, two negative parity states, $1^-$ and $3^-$, could be present in the spectrum.

\begin{figure}[t!]
\begin{center}
\includegraphics[width=9cm]{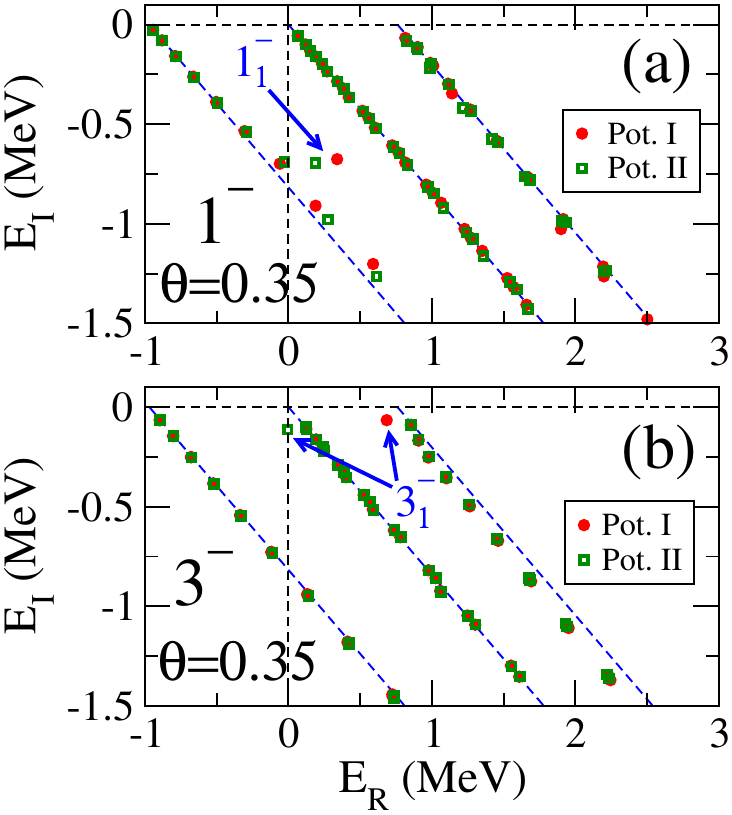}
\caption{The same as figure~\ref{fig2} for the (a) $1^-$ and (b) $3^-$ states.}
\label{sp13}
\end{center}
\end{figure}

Focusing again on potentials I and II, when no three-body force is employed, our calculations do not give rise to any 1$^-$ state, or at least they are not reachable with the complex scaling angle used in the calculations, $\theta$=0.35 rads.  However, when the three-body potential is used to introduce some additional attraction, a wide resonant state starts soon to show up. In figure~\ref{sp13}a we show the 1$^-$ discretized continuum spectrum obtained with $S_\mathrm{3b}=-6$ MeV and $\theta$=0.35 rads. We see that with both potentials, a resonance is now present at about 0.3 MeV above the three-body threshold and a width of about 1.4 MeV. The precise values are given in Table~\ref{tab2}.  Although the computed width agrees with the one of the tentative  experimental $1^-$ resonance, $1.3\pm0.5$ MeV \cite{til04}, the energy is clearly smaller than the experimental one, $1.25\pm0.20$ MeV. In principle this energy could be reached by making the three-body force less attractive, but in that case the computed width increases rapidly, and the resonance disappears in the $^6$He($0^+$)+$^2n$ continuum far before reaching the experimental energy.

\begin{table}[t!]
\caption{The same as Table~\ref{tab:gscomp} for the $1^-_1$  state with potentials I and II. }
\label{tab1-}
\begin{indented}
\item
\begin{tabular}{@{}lllll|lllll} \hline
\multicolumn{10}{c}{1$^-_1$ state }  \\ 
\multicolumn{5}{c}{$4n$($^2n$,$^2n$)+$\alpha$}  & \multicolumn{5}{c}{$^6$He($\alpha$,$^2n$)+$^2n$}  \\\hline
 $\ell_x$   & $\ell_y$ &$L$  &  \multicolumn{2}{c|}{weights} &  $\ell_x$ &$\ell_y$   &  $L$ &  \multicolumn{2}{c}{weights} \\ \cline{4-5} \cline{9-10}
   &    &   &    I &          II  &    &   &    &      I &  II  \\  
 0    & 1 &  1  &  100.0  &  97.3   &  0  & 1  & 1 & 45.1 & 49.1  \\
 2    & 1  &  1 &       0.0  &      1.2  &  1  & 0  & 1 & 35.1 & 33.2 \\
 2    & 3  &  1  &      0.0   &      1.5  &  1 &  2  & 1 &   1.6   &   1.9    \\
       &     &      &               &              &  2 & 1  & 1 &     13.4     &   10.3 \\
       &     &      &               &              &  2 & 3  & 1 &     2.2     &   2.5 \\
       &     &      &               &              &  3 & 2  & 1 &     2.6     &   3.0 \\
\hline
\end{tabular}
\end{indented}
\end{table}

Therefore, within our model, the computed $1^-$ state either has an
energy about 1 MeV below the experimental value and a width matching
experiment, or reproducing the measured energy with a width much
larger than the experimental value. In any case, it is worth noticing
that the component with a relative $p$-wave between the
$\alpha$-particle and the dineutron plays a very relevant role in this
case, contributing with about 35\% of the norm of the wave
function. This is seen in the right part of Table~\ref{tab1-}, where
we give the weight of the different components for the $1^-$ resonance
shown in figure~\ref{sp13}a.  Since the $\alpha$-dineutron $p$-wave
potential has not been adjusted to any $^6$He property, but simply has
been taken equal to the $\ell$=0 potential, a more careful treatment
of this interaction could modify the results.  For example,
reproduction of the $1^-$ state in $^8$He might be achieved by adjusting the
unknown $1^-$ in $^6$He.  These properties are connected in our model.
From the left part of the table we can see that this state is fully given by a single
component in the first Jacobi set, with the two dineutrons in a relative $s$-wave and the $\alpha$ in a $p$-wave relative to the $^2n$-$^2n$ center of
mass.

\begin{table}[t!]
\caption{The same as Table~\ref{tab:gscomp} for the $3^-_1$  state with potentials I and II. }
\label{tab3-}
\begin{indented}
\item
\begin{tabular}{@{}lllll|lllll} \hline
\multicolumn{10}{c}{3$^-_1$ state }  \\ 
\multicolumn{5}{c}{$4n$($^2n$,$^2n$)+$\alpha$}  & \multicolumn{5}{c}{$^6$He($\alpha$,$^2n$)+$^2n$}  \\\hline
 $\ell_x$   & $\ell_y$ &$L$  &  \multicolumn{2}{c|}{weights} &  $\ell_x$ &$\ell_y$   &  $L$ &  \multicolumn{2}{c}{weights} \\ \cline{4-5} \cline{9-10}
   &    &   &    I &          II  &    &   &    &      I &  II  \\  
 0    & 3 &  3  &  47.2    &  51.9 &  0  & 3  & 3 & 10.8 & 7.7  \\
 2    & 1 & 3   &   49.5   &   38.1  &  1  & 2  & 3 &  9.3 & 11.8 \\
 2   &   3& 3   &    3.3     &    2.2    &  2 &  1  & 3 &  74.9    &  74.6     \\
      &     &     &                &               &  2 & 3  & 3 &    1.2     &    0.2 \\ 
     &     &     &                &               &  3  & 0  & 3 &     1.3     &      2.0 \\
     &    &     &                 &               &  3 & 2  & 3 &     2.5     &      1.5 \\
\hline
\end{tabular}
\end{indented}
\end{table}

We close this section with the results for the $3^-$ states. The computed discretized continuum spectrum is shown in figure~\ref{sp13}b for potentials I and II. These results have
been obtained without inclusion of a three-body force. We can see that one resonance can be clearly seen for each of the two potentials, whose energies
and widths  are given in the last row of Table~\ref{tab2}. The computed energies are several MeV below the tentative experimental value given in reference~\cite{til04}. 
This means therefore that a quite repulsive three-body force should be needed in order to place the resonance at the experimental value. In any case, when we try to do this, 
we find that the $3^-$ resonance disappears rather fast in the continuum.

In Table~\ref{tab3-} we give the contribution of the components included in the calculation of the $3^-$ state (with no three-body force).  We see that this state is dominated 
by $^6$He in the $2^+$ state, and, contrary to the $1^-$ case, the $p$-wave $\alpha$-dineutron component plays a limited role (about 10\% of the norm).  Therefore, the 
uncertainty coming from the unknown $p$-wave $\alpha$-dineutron interaction is not that crucial in this case.

\subsection{Properties of excited $^8$He states}

The two best potentials are used to obtain excited states.  Two of
them, $0^+_2$ and $2^+_1$, are experimentally seen.  Two more states
are also known experimentally with the same tentative spin and parity,
$1^-_1$ and $3^-_1$, as found in this work.  The computed energies,
especially of the $3^-_1$ state, do not fully agree with the
experimental values. In any case, the computed odd parity energies are
less reliable, because an uncertainty from an unknown $p$-wave in
$^6$He($\alpha$+$^2n$) contributes significantly, mainly for the
$1^-_1$ state. We predict the existence of a $2^+_2$ state.  Energies
and radii are in fair agreement with the available data.  We did not
fine-tune the interactions to reproduce experiments more precisely.

In contrast to the ground state, the partial wave decomposition for
the second $0^+$ state reveals a content of about $73$\%
$^6$He($\alpha$+$^2n$) in the $2^+$ excited state.  The
average-distance density distribution for this excited state is
delicate, since it is found as a resonance after complex scaling.  The
structure reveals very well separated clusters, where the dineutrons
to a large extent are located on each side of the $\alpha$-particle.

The monopole transition between the two $0^+$-states is less precisely
defined, since $0^+_2$ is a continuum structure. The spurious
large-distance contribution from $^6$He($0^+_1$)$+^2n$ has to be
removed, but unfortunately difficult to differentiate for real
three-body energies, from the desired short-distance contributions of
$0^+_2$.  Still, we give an estimate, which within the inherent
uncertainties agree with the measured value.

The partial wave decompositions reveal predicted substructures.  The
$2^+_1$-state has about $36$\% and $58$\%, respectively in ground and
first excited $^6$He($\alpha$+$^2n$) states, whereas $2^+_2$ has about
$85$\% in the $^6$He excited $2^+$ state.  The $1^-_1$-state has all
strength in the $s$-wave between the two dineutrons. Translation to
the other Jacobi coordinate showing $^6$He($\alpha$+$^2n$) with an
attached dineutron is then equally sharing $s$ and $p$ waves between
$\alpha$ and $^2n$. The $3^-_1$-state is sharing almost equally the
$s$ and $d$-waves between dineutrons, but arising from different
couplings to the $\alpha$-particle.  The translation to the other
Jacobi coordinate gives about $75$\% of the $^6$He $2^+$-state.  The
estimates of the odd parity structures are more uncertain, because
they depend on the unknown $p$-wave of $^6$He.

The structure of the four valence neutrons in $^8$He is tested by
pushing the $\alpha$-core gently away by a fast moving projectile and
measuring the missing-mass spectrum.  We believe this is a relatively
sensitive structure quantity, which in the present calculation with
dineutrons agrees with the measurement.

\section{$^7$He as an $\alpha$+$^2n$+$n$ system}

Focusing now on the $^7$He nucleus, our aim here is to investigate how the $\alpha$+$^2n$+$n$ structure is able to describe its main properties. Following the results 
obtained for $^8$He, where we have seen that a very weak, or even zero, $^2n$-$^2n$ interaction is more appropriate to describe the nucleus, we assume here that the same
applies  for the $^2n$-$n$ interaction.  We have then made the calculations for $^7$He taking the    $^2n$-$n$  potential equal to zero.

\begin{figure}[t!]
\begin{center}
\includegraphics[width=9cm]{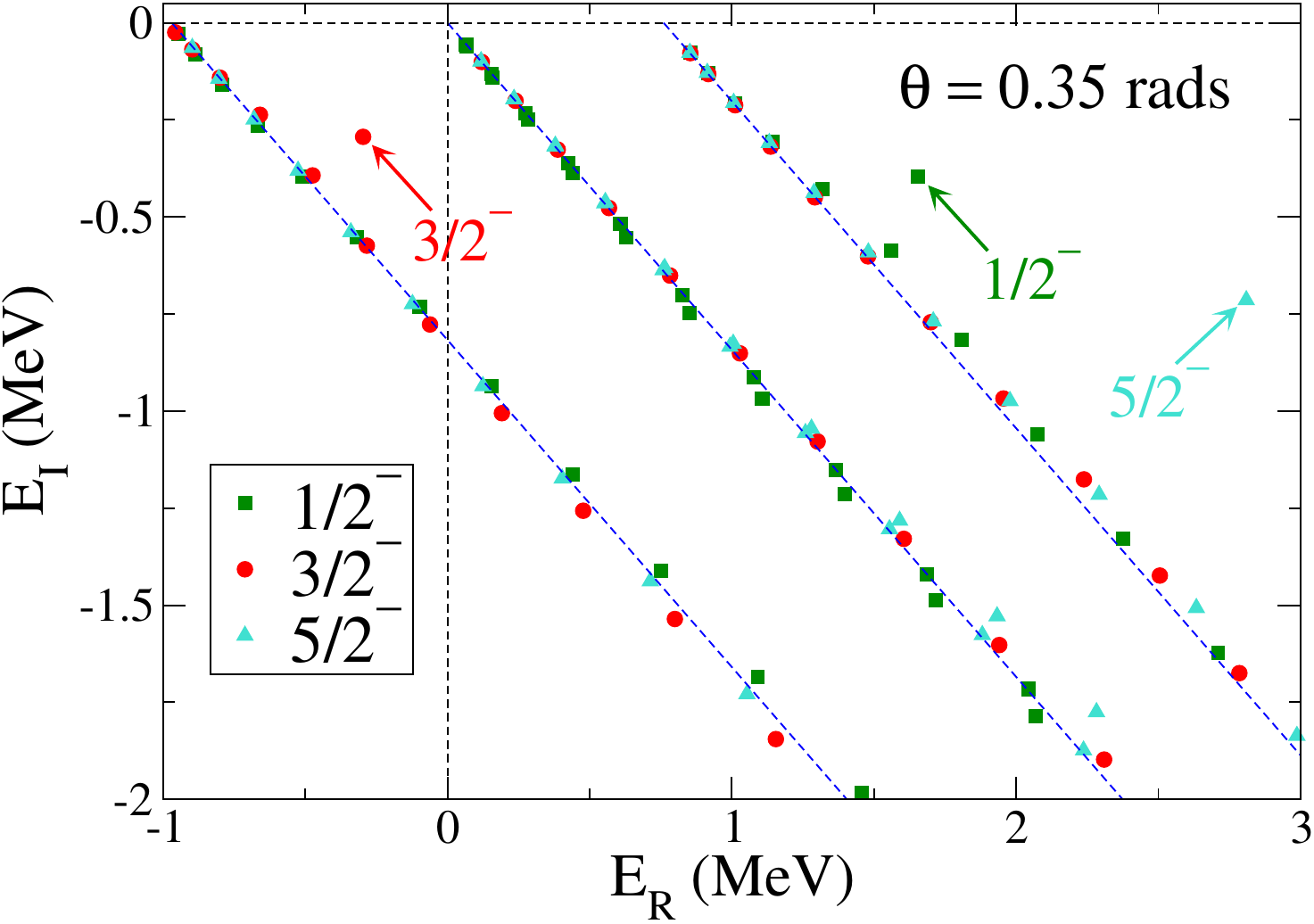}
\caption{Complex scaled energy spectrum of the  $\frac{1}{2}^-$, $\frac{3}{2}^-$, and $\frac{5}{2}^-$ states in $^7$He.}
\label{spec7He}
\end{center}
\end{figure}

\begin{table}
\caption{Energies, $E_R$, and widths, $\Gamma_R$, of the computed $^7$He states.  
The energies (with respect to the  three-body threshold) and the widths are given in MeV. 
$S_\mathrm{3b}$ gives the strength, in MeV, of the Gaussian three-body force in equation~(\ref{gau3b}). 
The experimental data have been taken from \cite{til02}. }
\label{7He}
\begin{indented}
\item 
\begin{tabular}{@{}l|lll|ll} \hline
$J^\pi$ &   $E_R$ &  $\Gamma_R$ &  $S_\mathrm{3b}$ & $E_R$ &  $\Gamma_R$ \\ \hline
$\frac{3}{2}^-$  &  $-0.52$ & 0.35 &  $-0.84$ &  $-0.529$  &  0.15$\pm$0.02 \\
$\frac{1}{2}^-$  & 1.65  & 0.79  & 0.0  &  &       \\
$\frac{5}{2}^-$  & 2.39  &  0.94 &  $-2.0$ & 2.39$\pm$0.09 &  1.99$\pm$0.17    \\ \hline
\end{tabular}
\end{indented}
\end{table}

When this is done we obtain for $^7$He the discretized continuum spectrum shown in figure~\ref{spec7He} for the $\frac{1}{2}^-$,  $\frac{3}{2}^-$, and $\frac{5}{2}^-$
states. We can immediately see that the ground state is given by a resonant $\frac{3}{2}^-$ state at $-0.30$ MeV with respect to  the three-neutron threshold, or, in other words,
0.67 MeV above the $^6$He($0^+$)+$n$ threshold. This energy is close to the experimental value, $-0.529$ MeV given in reference~\cite{til02}. A small attractive
three-body force is then sufficient to fit the experiment ($S_\mathrm{3b}=-0.84$ MeV in equation~(\ref{gau3b})). The resulting
energy and width are quoted in the first row of Table~\ref{7He}, where we can see that the width of the computed resonance is about a factor of two bigger than the 
experimental one.

\begin{table}[t!]
\caption{For the $\frac{3}{2}^-$  and $\frac{1}{2}^-$ states in $^7$He, and for each of the  three different Jacobi sets, contribution to the norm of the wave function of the components 
included in the calculation. }
\label{tab32-}
\begin{indented}
\item 
\begin{tabular}{@{}lllll|lllll|lllll} \hline
\multicolumn{15}{c}{$\frac{3}{2}^-$ and $\frac{1}{2}^-$ states }  \\ 
\multicolumn{5}{c}{$3n$($^2n$,$n$)+$\alpha$}  & \multicolumn{5}{c|}{$^5$He($\alpha$,$n$)+$^2n$}  &  \multicolumn{5}{c}{$^6$He($\alpha$,$^2n$)+$n$}  \\ \hline
 $\ell_x$   & $\ell_y$ &$L$  & \multicolumn{2}{c|}{weights} &  $\ell_x$ &$\ell_y$   &  $L$ &  \multicolumn{2}{c|}{weights}   & $\ell_x$ &$\ell_y$   &  $L$ &  \multicolumn{2}{c}{weights}  \\ 
 \cline{4-5}  \cline{9-10}   \cline{14-15}
           &    &      &   $\frac{3}{2}^-$  &  $\frac{1}{2}^-$  &      &    &     &   $\frac{3}{2}^-$      &  $\frac{1}{2}^-$   &       &   &   &    $\frac{3}{2}^-$      &  $\frac{1}{2}^-$ \\
   0    & 1 &  1  &  35.6  & 48.6   &  0  & 1  & 1 & 1.3    & 16.2 &  0 &   1 & 1  &   91.8  &  1.7 \\
   1    & 0 &  1  &  64.4   &21.9  &  1  & 0  & 1 & 98.7   & 0.0    & 1  & 0 & 1 &   3.8     &   7.3\\
   1    & 2 &  1  &   0.0   &   29.5 &  1  & 2  & 1 &  0.0    & 82.7 & 1  & 2 & 1 &   2.7     &    0.0\\
   2    & 1 &  1  &   0.0   & 0.0  &  2  & 1  & 1 &  0.0       &1.1     & 2  & 1 & 1 &   1.1   &     88.0\\
   2    & 3 &  1  &   0.0   & 0.0  &  2  & 3  & 1 &  0.0      & 0.0     & 2  & 3 & 1 &   0.6   &   0.0  \\
   3    & 2 &  1  &   0.0   & 0.0  &  3  & 2  & 1 &  0.0      & 0.0     & 3  & 2 & 1 &   0.0   &   3.0\\
\hline
\end{tabular}
\end{indented}
\end{table}

In figure~\ref{spec7He} we also see that the first excited state corresponds to spin and parity $\frac{1}{2}^-$, which appears at an energy 1.65 MeV above the three-neutron
threshold, with a width of about 0.8 MeV. So far, there has not been experimental  evidence of such an excited state. 

It is important to note here that the components entering
in the description of the $\frac{3}{2}^-$ and $\frac{1}{2}^-$ states are the same, since they both result from the coupling of the total orbital angular momentum $L=1$ and the total
spin $S=\frac{1}{2}$. The only difference between the two states comes from the different weight of the components. These weights are shown in Table~\ref{tab32-} for the two
states and the three possible Jacobi sets. The different structure of the $\frac{3}{2}^-$ and $\frac{1}{2}^-$ states can be seen in the central part of the table, which indicates that both
states are mainly given by $^5$He in a $p$-state, but with the dineutron in a relative $s$-wave in the $\frac{3}{2}^-$ case, or in a relative $d$-wave in the $\frac{1}{2}^-$ case.
A similar difference can be seen in the Jacobi set shown in the right part of the table, which shows that the structure of these two states can also be understood as the neutron
in relative $p$-wave with respect to $^6$He, but whereas $^6$He is in a $0^+$ state for the $\frac{3}{2}^-$ state, it is in a $2^+$ state in the $\frac{1}{2}^-$ state.

\begin{table}[t!]
\caption{The same as Table~\ref{tab32-} for the $\frac{5}{2}^-$ state. }
\label{tab52-}
\begin{indented}
\item
\begin{tabular}{@{}llll|llll|llll} \hline
\multicolumn{12}{c}{$\frac{5}{2}^-$ state }  \\ 
\multicolumn{4}{c}{$3n$($^2n$,$n$)+$\alpha$}  & \multicolumn{4}{c|}{$^5$He($\alpha$,$n$)+$^2n$}  &  \multicolumn{4}{c}{$^6$He($\alpha$,$^2n$)+$n$}  \\ \hline
 $\ell_x$   & $\ell_y$ &$L$  &  weight &  $\ell_x$ &$\ell_y$   &  $L$ &  weight   & $\ell_x$ &$\ell_y$   &  $L$ &  weight  \\ 
   1    & 2 &  2  &  68.0     &  1  & 2  & 2 & 90.4  &  1 &   2 & 2  &   0.2\\
   2    & 1 &  2  &  22.1     &  2  & 1  & 2 & 8.1    & 2  & 1 & 2 &   99.8 \\
   2    & 3 &  2  &   8.7      &  2  & 3  & 2 &  1.5    & 2  & 3 & 2 &   0.0 \\
   3    & 2 &  2  &   1.2      &  3  & 2  & 2 &  0.0    & 3  & 2 & 2 &   0.0 \\
\hline
\end{tabular}
\end{indented}
\end{table}

Finally, in figure~\ref{spec7He} we see as well that the calculations give rise to an excited $\frac{5}{2}^-$ state with energy 2.81 MeV above the three-neutron threshold and a 
width of 1.4 MeV. In reference~\cite{til02} a resonant state at 2.4 MeV above the three-neutron threshold and width of about 2.0 MeV is reported, and the tentative spin 
and parity of $\frac{5}{2}^-$ is assigned.  It is then reasonable to assume that our computed state actually corresponds to this experimentally known resonance. When
the three-body potential is used to fit the experimental value (which is done with $S_\mathrm{3b}$=$-2.0$ MeV), we get the result given in the last row
of Table~\ref{7He}. We see that the width of the computed state is then about 1 MeV, roughly half the size of the experimental value.

For completeness we close this section giving in Table~\ref{tab52-} the contribution to the $\frac{5}{2}^-$ wave function of the different components included in the calculation.  
We see that the wave function is to a very large extent given by $(\ell_x=1, \ell_y=2)$ component in the second Jacobi set ($^5$He in a $p$-state and $^2n$ in a relative
$d$-wave),  or the $(\ell_x=2, \ell_y=1)$ component in the third Jacobi set ($^6$He in a $d$-state and $n$ in a relative $p$-wave). These two are also the components
giving rise to the $\frac{1}{2}^-$ state (see Table~\ref{tab32-}), but with the crucial difference that whereas in the $\frac{1}{2}^-$ case $\ell_x$ and $\ell_y$ couple to $L=1$,
in the $\frac{5}{2}^-$ case they couple to $L=2$.

To close this section let us mention that, as done in the $^8$He case, the calculations could have been done assuming a small  $^2n$-$n$ interaction such that three-body potential is not needed. For the $\frac{3}{2}^-$ state this is achieved with an attractive Gaussian potential with strength $-6.3$ MeV and range $1.5$ fm. This
gives rise to a resonance energy of $-0.52$ MeV and a width of 0.32 MeV, very much the same as quoted in  Table~\ref{7He}. Furthermore, the partial wave content
is also very similar to the one given in Table~\ref{tab32-}. The computed content of the dominating $(0,1,1)$ and $(1,0,1)$ components is now $(36.5,63.5)$, $(1.0,99.0)$ and $(91.8,4.4)$ in the first, second, and third Jacobi sets, respectively.  As one can see, the change is minimal. For the other two states, $\frac{1}{2}^-$ and $\frac{5}{2}^-$,  the variation in the partial wave content  compared to the results given in Tables~\ref{tab32-} and \ref{tab52-} is also very small, not relevant either.

\section{Summary and conclusions}

We have investigated two structures with three and four neutrons outside an
$\alpha$-particle core. In the shell model, all neutrons would be in
the $p_{3/2}$ state, but it may be advantageous to exploit other degrees of
freedom. Since neutrons are fermions, the most obvious is to form
pairs of clusters, i.e. the entities $n$ and $^2n$.  This attempt is
also suggested by the hyper-radial prediction that only two- and
three-body clusters are likely to exist.  Thus, we assume the
structures $^7$He($\alpha$+$^2n$+$n$) and $^8$He($\alpha$+$^2n$+$^2n$)
and perform the corresponding three-body calculations.  The dineutron,
$^2n$, is treated as a spin-zero particle without intrinsic structure.

The known subsystems, $^5$He and $^6$He, are used to constrain the
two-body interactions, $\alpha$-$^2n$ and $\alpha$-$n$.  This
incorporates effects of possible intrinsic $^2n$ structure.  The other
two-body interactions, $^2n$-$^2n$ and $^2n$-$n$, are required to
reproduce the energies of the three-body (ground) states. They turn
out to be rather weak and almost compatible with zero. The adiabatic
hyperspherical expansion method supplemented by complex scaling is
applied to obtain three-body solutions for different angular momenta
and parities. If necessary, a three-body potential is added to match
precisely the ground state energy.

The computations are in fair agreement with the available data, first
of all the energy spectra, where the existence of a second $2^+$ in
$^8$He and one $\frac{1}{2}^-$ in $^7$He are predicted.  In addition,
the average-distance density distributions for both $0^+$ states in
$^8$He show non-overlapping three-body constituents.  The missing mass
spectrum of the four neutrons after removal of the $\alpha$-particle
as well as the monopole transition between the two $0^+$-states are
within error bars consistent with measurements.

The calculated partial wave decompositions describe the structures of
the states.  The realistic potentials reveal that the $0^+$ ground
state of $^8$He has a content of $97$\% of $0^+$ ground state of
$^6$He.  In contrast, the second $0^+$ state has a content of about
$73$\% of $^6$He in the $2^+$ excited state. Both with $^6$He
surrounded by a dineutron $^2n$.  The $2^+_1$-state has about $36$\%
and $58$\%, respectively in ground and first excited
$^6$He($\alpha$+$^2n$) states, whereas $2^+_2$ has about $85$\% in the
$^6$He excited $2^+$ state.  The $1^-_1$-state has all strength in the
$s$-wave between the two dineutrons.  The $3^-_1$-state has about
$75$\% of the $^6$He $2^+$-state.

The $\frac{3}{2}^-$ ground state of $^7$He can be understood either as
$99$\% $^5$He surrounded by an $s$-wave dineutron, or as $92$\% $^6$He
surrounded by a $p$-wave neutron.  The dominating part of the
$\frac{1}{2}^-$ state can also be understood either as $83$\% $^5$He
surrounded by a $d$-wave dineutron, or as $88$\% $^6$He*($2^+$)
surrounded by a $p$-wave neutron.  The highest found resonance,
$\frac{5}{2}^-$, is described either as $90$\% $^5$He surrounded by an
$d$-wave dineutron, or as $100$\% $^6$He*($2^+$) surrounded by a
$p$-wave neutron.  The different descriptions of the same state are
the consequence of the reference to different bases.

In conclusion, the dominating components in the structures of $^7$He
and $^8$He can be described as three-body systems consisting of an
$\alpha$-particle combined with one neutron and one dineutron and two
dineutrons, respectively.  In this work, we derived contributions of
substructures for each three-body state, which expressed in different
bases appears different, although it is the same state.  We see
connections between interactions and structures of nuclei in the
isotopic chain.  Specifically, a measurement of $^8$He($1^-$) would
provide information about an unknown $^6$He($1^-$), and the $^8$He
spectrum show that the effective interactions between $^2n$-$^2n$ and
$^2n$-$n$ are very weak.  The perspective of this work is that a
number of other neutron dripline nuclei can be understood as specific
two or three-body clusters. The main uncertainty in such descriptions
would arise from nucleons arranged in several different (coupled)
clusters.

\ack This work has been partially supported by: Grant PID2022-136992NB-I00 funded by MCIN/AEI/10.13039/501100011033.

\end{document}